\documentclass[10pt, conference, compsocconf]{IEEEtran}

\usepackage{graphicx}
\usepackage{times}
\usepackage{helvet}
\usepackage{courier}
\usepackage{booktabs}
\usepackage[numbers,sort]{natbib}
\usepackage{epsfig}
\usepackage{amssymb}
\usepackage{amsmath}
\usepackage{amsfonts}
\usepackage{breqn}
\usepackage{multirow}
\usepackage{array}
\usepackage{subcaption}
\usepackage{mathtools}
\usepackage{caption}
\usepackage{url}
\usepackage{float}
\usepackage{balance}
\begin{document}

\newcommand{\eq}[1]{Eq.~(\ref{#1})}

\newcommand{\xhdr}[1]{\vspace{1mm}\noindent{{\bf #1}}}
\newcommand{\md}{Monomer}
\newcommand{\modelnamelong}{Mixtures of Non-Metric Embeddings for Recommendation}

\title{Learning Compatibility Across Categories for Heterogeneous Item Recommendation}


\author{\IEEEauthorblockN{Ruining He, Charles Packer, Julian McAuley}
\IEEEauthorblockA{Department of Computer Science and Engineering\\
University of California, San Diego\\
Email: \{r4he, cpacker, jmcauley\}@cs.ucsd.edu}
}

\maketitle
\begin{abstract}
Identifying relationships between items is a key task of an online recommender system, in order to help users discover items that are functionally complementary or visually compatible. In domains like clothing recommendation, this task is particularly challenging since a successful system should be capable of handling a large corpus of items, a huge amount of relationships among them, as well as the high-dimensional and semantically complicated features involved. Furthermore, the human notion of ``compatibility'' to capture goes beyond mere similarity: For two items to be compatible---whether jeans and a t-shirt, or a laptop and a charger---they should be similar in some ways, but systematically different in others.

In this paper we propose a novel method, \emph{\md{}}, to learn complicated and heterogeneous relationships between items in product recommendation settings. Recently, scalable methods have been developed that address this task by learning similarity metrics on top of the content of the products involved. 
Here our method relaxes the \emph{metricity} assumption inherent in previous work and models multiple localized notions of `relatedness,' so as to uncover ways in which related items should be systematically similar, and systematically different. Quantitatively, we show that our system achieves state-of-the-art performance on large-scale compatibility prediction tasks, especially in cases where there is substantial heterogeneity between related items. Qualitatively, we demonstrate that richer notions of compatibility can be learned that go beyond similarity, and that our model can make effective recommendations of heterogeneous content.
\end{abstract}


\IEEEpeerreviewmaketitle

\section{Introduction}
Identifying and understanding relationships between items is a key component of any modern recommender system. Knowing which items are `similar,' or which otherwise may be substitutable or complementary, is key to building systems that can understand a user's context, recommend alternative items from the same style \cite{Etsy}, or generate bundles of items that are compatible \cite{Linden03,ebay,VisualSIGIR}.

Typically, identifying these relationships means defining (or otherwise learning from training data) an appropriate distance or similarity measure between items. This is appropriate when the goal is to learn some notion of `equivalence' between items, e.g.~in order to recommend an item that may be a natural alternative to the one currently being considered.
However, identifying such a similarity measure may be insufficient when there is substantial \emph{heterogeneity} between the items being considered. For example, the characteristics that make clothing items, electronic components, or even romantic partners compatible exhibit substantial heterogeneity: for a pair of such items to be compatible they should be systematically similar in some ways, but systematically different in others.

Recently, a line of work has aimed to model such heterogeneous relationships, e.g.~to model co-purchasing behavior between products based on their visual appearance or textual descriptions~\cite{VisualSIGIR,SiameseICCV,McAPanLes15}. In spite of the substantial heterogeneity in the data used for training (a large dataset of co-purchase `dyads' from \emph{Amazon}) and the complexity of the models used, these works ultimately follow an established metric-learning paradigm: (1) Collect a large dataset of related (and unrelated) items; (2) Propose a parameterized similarity function; and (3) Train the parameterized function such that related items are more similar than non-related items. Such metric-learning approaches can be incredibly flexible and powerful, and have been used to identify similarities between items ranging from music \cite{slaney2008learning} to members of the same tribe \cite{Der12}. 
Such methods work to some extent even in the presence of heterogeneity, since they learn to `ignore' dimensions where similarity should not be preserved. But we argue that ignoring such dimensions discards valuable information that ought to be used for prediction and recommendation.

In this paper, we propose new models and algorithms to identify relationships between items in product recommendation settings. In particular, we relax the metricity assumption present in recent work, by proposing more flexible notions of `relatedness' while maintaining the same levels of speed and scalability. Specifically, we hope to overcome the following limitations of previous work:
\begin{itemize}
\item The similarity measures learned by previous approaches ultimately project categories as clusters into a metric space (albeit potentially via a complex embedding), since an item is inherently more similar to those from the same category than others (as we show later in Figure \ref{fig:lmtembed}). 
This means that cross-category recommendations can only be made by exploiting an explicit category tree (e.g.~`find the shoes nearest to these jeans'). Not only do such approaches require explicit category labels, but they are also subject to any noise or deficiencies in the category data. Our method can make cross-category recommendations without any dependence on the presence (or quality) of explicit category information.

\begin{figure}[!t] 
\centering
\includegraphics[width=\columnwidth]{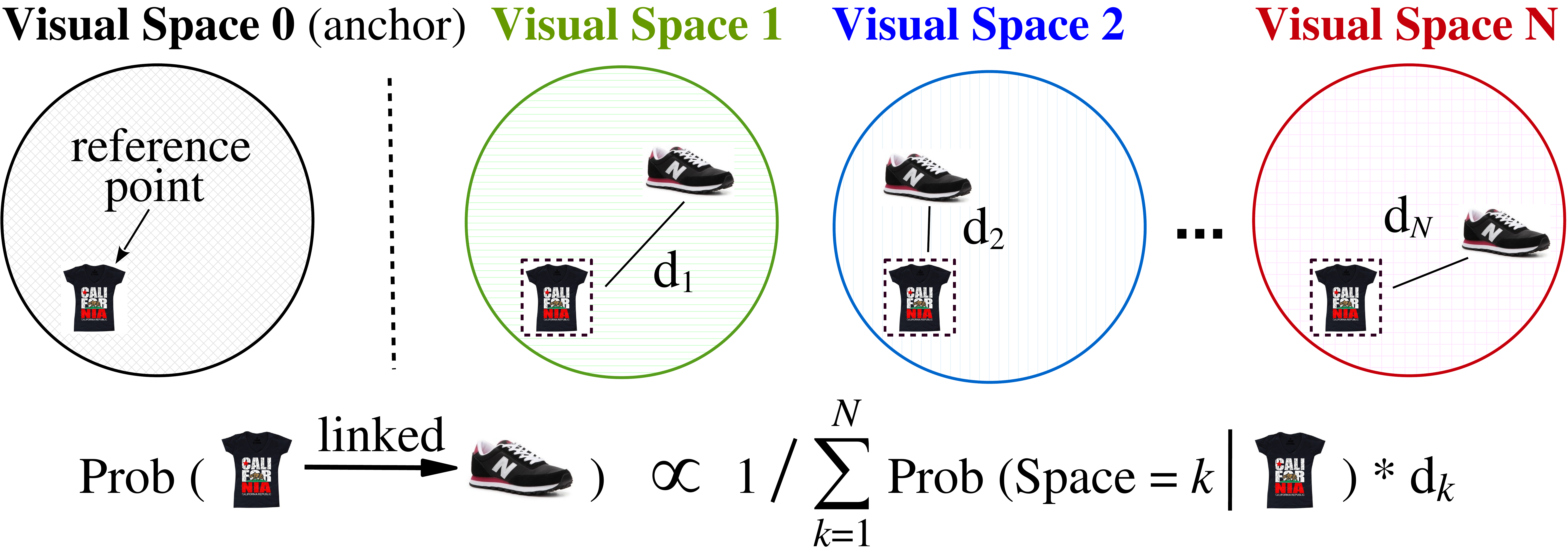}
\caption{Illustration of the high-level ideas of \emph{\md{}}. The query item (a t-shirt) is embedded into visual space 0 (the anchor space) whose position we superimpose into the other spaces. The potential match (a shoe) is embedded to $N$ visual spaces and within each of them Euclidean distance between the pair is computed. Finally, the mixtures-of-experts framework is adopted to model the relative importance of the different components w.r.t.~the given query. We show this on a real example in Figure \ref{fig:todo}.}
\label{fig:idea}
\end{figure}

\item Other assumptions made by metric learning approaches are also too strict for recommendation: an item is not necessarily compatible with itself (identity), 
nor are the types of relationships we want to learn necessarily symmetric (e.g.~a spare battery is a good add-on item for a laptop, but not vice-versa). 
Other assumptions hidden in previous approaches (such as transitivity) may also be too strict, e.g.~an iPhone is dissimilar from a Surface, though both are related to an iPad.
Our approach is flexible enough to capture such complex and non-metric relationships.
\item Previous approaches learned 
a single `global' (albeit complex) notion of relatedness, neglecting any `local' notions that could be equally important. 
In contrast, we capture multiple (and possibly competing) notions of `relatedness' simultaneously. This is also key to generating \emph{diverse sets} of recommendations. E.g.~a shirt may be compatible with (a) a similar shirt from a different brand, (b) a similar shirt with a different color, (c) a complementary pair of pants, or (d) a complementary pair of shoes. By learning `relatedness' as a mixture of multiple competing notions, we can handle diverse sets of recommendations naturally.
\end{itemize}

We propose a novel method, \emph{\modelnamelong{}}, or \emph{\md{}} for short, that addresses the above limitations. We demonstrate our idea in Figure~\ref{fig:idea} (we later show an example on real data in Figure~\ref{fig:todo}). Here we embed the first item 
$x$ (the query) into one space (the `anchor space'), and embed its potential match 
$y$ into a series of $N$ additional spaces. 
Now, the relatedness between $x$ and $y$ is measured in terms of multiple notions, each captured by one of the $N$ spaces involved.
Furthermore, the $N$ spaces are \emph{weighted} according to a mixtures-of-experts type framework, determining 
to what extent each of the $N$ embeddings is `relevant' to a particular query.

Note in particular that the method described in Figure~\ref{fig:idea} can learn \emph{non-metric} relationships since we are measuring the distance between two \emph{different} embeddings.
The learned relationships are not necessarily symmetric, nor do identity and transitivity necessarily hold; on the other hand the model is flexible enough such that a metric embedding \emph{could} be learned if that was what the data supported.

For clarity, our contributions are summarized as follows:
\begin{enumerate}
\item We propose a new scalable method, \emph{\md}, for heterogeneous item-to-item recommendation. The presented mixtures-of-embeddings framework allows it to learn \emph{non-metric} relationships, thereby overcoming multiple limitations present in existing work. 
\item We demonstrate quantitatively that \emph{\md{}} is effective at learning notions of `relatedness' from heterogeneous dyads of co-purchases from \emph{Amazon}, and in particular that it does so more accurately than recent approaches based on metric/similarity learning.
\item We qualitatively show that \emph{\md{}} can effectively learn multiple, semantically complex notions of `relatedness,' and that these can be useful to generate rich, heterogeneous, and diverse sets of recommendations.
\end{enumerate}

Code and data can be found at \url{https://sites.google.com/a/eng.ucsd.edu/ruining-he/}.

\section{Related Work}

The most closely related branches of work to ours are (1) Those that deal with the item-to-item recommendation problem, e.g.~systems that generate recommendations by modeling relationships between items; and (2) Works that deal with metric (or relationship) learning in general, whether or not for recommendation.

\xhdr{Item-to-Item Recommendation.} Identifying relationships among items is a fundamental part of many real-world recommender systems, e.g.~to generate recommendations of the form `people who viewed $x$ also viewed $y$' on \emph{Amazon}. Such methods may be based on collaborative filtering, e.g.~counting the overlap between users who have clicked on / bought both items, as in \emph{Amazon's} own solution \cite{Linden03}. Latent-factor approaches aim to model user-item relationships in terms of low-dimensional factors, such that `similar' items are those with close embeddings. See (e.g.) \cite{adomavicius2005toward,korenSurvey} for surveys, and \cite{Linden03}, \cite{Etsy}, and \cite{ebay} for specific systems that make use of \emph{Amazon}, \emph{Etsy}, and \emph{Ebay} data.

Of more interest to us are systems that predict item-to-item relationships based on the \emph{content} (e.g.~images/text/metadata) of the items themselves. Various systems have been proposed to address specific settings, e.g.~to identify relationships between members of `urban tribes' \cite{murillo2012urban}, tweets \cite{learning1}, text \cite{blockLDA,chang2009relational}, or music \cite{slaney2008learning}. Several methods have also been used to model visual data \cite{carneiro2007supervised,doersch2012what,amazonJaiwei,shrivastava-sa11,kernel1}, though typically in settings where the metric assumption is well-founded (e.g.~similar image retrieval).

Our work follows recent examples that aim to model co-purchase and co-browsing relationships, using a recently introduced dataset from \emph{Amazon} \cite{VisualSIGIR,SiameseICCV,McAPanLes15}. While we extend (and compare quantitatively against) such work, our main contribution here is that we substantially relax the model assumptions to allow for more complex relationships than mere similarity between items.

Also of interest are a few works that model clothing data, particularly in recommendation settings, e.g.~a few recent works that aim to capture some notion of `style' include \cite{Di:2013,streetFashion,gettingTheLook,magicCloset,yamaguchi2013paper,HeMcA16a}. However the specific tasks considered there are quite different from the item-to-item recommendation task in which we are interested.

\xhdr{Metric (and non-metric) Learning.} Outside of the recommendation scenarios considered here, learning the features that describe relationships between objects is a vast topic. Typically, one is given some collection of putative relationships between items (i.e., a training set), and the goal is then to identify a (parameterized) function that can be tuned to fit these relationships, i.e.,~to assign observed relationships a higher likelihood or score than non-relationships. State-of-the-art methods identify hidden variables or factors that describe relationships among items \cite{Der12,Tor07}, e.g.~by factorizing the matrix of links between items \cite{menon}. Again, the main contributions we hope to make over such approaches are (1) to relax the assumption of metricity, and (2) to allow for multiple notions of `relatedness' to compete and interact. While a few approaches have recently been proposed to learn non-metric relationships (e.g.~\cite{sca13}), we are unaware of any that allow for the scale of the data (thousands of features, millions of items and relationships) that we consider.

\section{The model} \label{sec:model}

Formally, we are given a dataset $\mathcal{D}$ comprising a large corpus of objects (or `items') and the pairwise relationships $\mathcal{R}$ between items from different subcategories, i.e., if $(x,y) \in \mathcal{R}$ then (1) item $x$ and $y$ are related, and (2) $x$ and $y$ are not from the same subcategory (e.g. a shirt and a matching pair of pants). We choose such cross-category recommendations to highlight the ability of our model to generate recommendations between heterogeneous pairs of items. This matches the training instance selection approach from \cite{SiameseICCV}. Additionally, a high-dimensional feature vector $f_x$ associated with each item $x$ is also provided (encoding e.g.~its image or the text of its reviews). We seek a scalable method to model such relationships with a set of parameterized transform functions $d(x,y)$ such that related objects ($(x,y) \in \mathcal{R}$) are assigned higher probabilities 
than non-related ones ($(x,y) \notin \mathcal{R}$).
Notation used throughout the paper is summarized in Table \ref{tab:notation}.

\begin{table}
\centering
\renewcommand{\tabcolsep}{2pt}
\caption{Notation \label{tab:notation}}
\begin{tabular}{lp{0.7\linewidth}} \toprule
Notation & Explanation\\ \midrule
$\mathcal{R}$ & relationships between a corpus of items\\
$(x,y)$ & an (ordered) item pair\\
$P((x,y)\in\mathcal{R})$ & probability that $x$ and $y$ are related\\
$F$ & dimensionality of the feature vector\\
$f_x, f_y$ & feature vectors of $x$ and $y$ respectively ($F \times 1$)\\
$K$ & rank of the Mahalanobis embedding matrix\\
$\mathbf{M}$ & Mahalanobis matrix ($F \times F$)\\
$\mathbf{E}$ & low-rank Mahalanobis matrix ($F \times K$)\\
$N$ & no. of Mahalanobis embeddings (i.e., learners)\\
$k$ & index of the $N$ learners ($k \in \{1,2,...,N\}$)\\
$\mathbf{E}_k$ & the $k$-th low-rank Mahalanobis matrix ($F \times K$)\\
$\mathbf{U}_k$ & parameter vector corresponding to learner $k$\\
$d_k(x,y)$ & distance from $x$ to $y$ predicted by learner $k$\\
$P(k|(x,y))$ & the `confidence' associated with learner $k$\\
$d(x,y)$ & distance from $x$ to $y$ (can be directed)\\
$\sigma_c(z)$ & shifted sigmoid function ($1/(1+\text{exp}(-z - c))$)\\

\bottomrule
\end{tabular}
\renewcommand{\tabcolsep}{6pt}
\end{table}

\subsection{Preliminaries}
\xhdr{Visual Features.} In this paper, we mainly consider the case of using high-level visual features for relationship prediction. This is particularly useful for clothing recommendation (for example), a natural domain in which learning heterogeneous relationships between items across categories is particularly important.

Our visual features are extracted from a deep convolutional neural network pre-trained on 1.2 million ImageNet (ILSVRC2010) images. In particular, we used the Caffe reference model \cite{CAFFE}, which has 5 convolutional layers followed by 3 fully-connected layers,
to extract $F = 4096$ dimensional visual features from the second fully-connected layer (i.e., FC7).

Note however that our proposed method is agnostic to the type of features used, and as we show later can handle other types of features (e.g.~text) in order to address more general settings.

\xhdr{Mahalanobis Transform.}
In order to model subtle notions like 
`compatibility'
upon the raw visual features, we need expressive transformations that are capable of relating feature dimensions to explain the relationships between pairs of items. To this end, we follow the approach from \cite{VisualSIGIR}: 
there, a \textit{Mahalanobis Distance} is used to measure the distance (or `dissimilarity') between items within the feature space according to the knowledge of how different feature dimensions relate to each other. Let $\mathbf{M}$ denote the matrix that parameterizes the Mahalanobis Distance, then the distance between an item pair $(x, y)$ is defined by
\begin{equation}
d_{\mathbf{M}}(x, y) = (f_x - f_y)^T \mathbf{M} (f_x - f_y),
\label{eq:mahal_basic}
\end{equation}
where $f_x$ and $f_y$ are the features vectors of $x$ and $y$ respectively. Although such an approach defines a distance function (and therefore suffers from the issues we are hoping to address), we use this method as a building block and ultimately relax its limitations.

\xhdr{Mixtures-of-Experts.}
Mixtures of experts (MoEs) are a classical machine learning method to aggregate the predictions of a set of `weak' learners, known as experts \cite{jacobs1991adaptive}. What is particularly elegant about this approach is that it allows each learner to `focus' on classifying instances about which it is relevant (i.e., expert), without being penalized for making misclassifications elsewhere.

For regression tasks such as the one we consider, each learner (denoted by $l$) outputs a prediction value $\mathit{Pred}_l(X)$ for the given input $X$. These predictions are then aggregated to generate the final prediction by associating weighted `confidence' scores with each learner. Here we are interested in probabilistically modeling such confidences to be proportional to the expertise of the learners:
\begin{equation}
\underbrace{\mathit{Pred}(X)}_{\mathclap{\text{final prediction}}} = \sum_l \overbrace{P(l|X)}^{\mathclap{\text{confidence in $l$'s expertise}}} \cdot \underbrace{\mathit{Pred}_l(X)}_{\mathclap{\text{$l$'s prediction}}}.
\end{equation}
In our model, each `expert' shall correspond to a single notion of `relatedness' between items. Thus, for a given pair of items that are potentially related, we can determine (a) which notions of relatedness are relevant for these items ($P(l|X)$); and (b) whether or not they are related according to that notion ($\mathit{Pred}_l(X)$). These two functions are learned jointly, such that the model automatically uncovers multiple notions of `relatedness' simultaneously.

\subsection{Model Specifics}
First we describe how Mahalanobis transforms have previously been applied to this task, and can be used as a building block for this task, before describing our proposed non-metric method.

\subsubsection{Low-rank Mahalanobis Metric}
Considering the high dimensionality of the visual features we are modeling (feature dimension $F$ = 4096 in our case), learning a full rank positive semi-definite matrix $\mathbf{M}$ as in \eq{eq:mahal_basic} is neither computationally tractable for existing solvers nor practical given the size of the dataset.

Recently it was shown in \cite{VisualSIGIR} that a low-rank approximation of a Mahalanobis matrix works very well on visual datasets for the tasks considered in this paper. Specifically, the $F \times F$ Mahalanobis matrix is approximated by $\mathbf{M} \approx \mathbf{E} \mathbf{E^T}$, where $\mathbf{E}$ is an $F \times K$ matrix and $K \ll F$. Then the distance between a pair $(x,y)$ is calculated by
\begin{equation} \label{eq:sigir}
d_{\mathbf{E}}(x, y) = (f_x - f_y)^T \mathbf{E} \mathbf{E}^T (f_x - f_y) = ||\mathbf{E}^T f_x - \mathbf{E}^T f_y||_2^2.
\end{equation}
This can be viewed as \emph{embedding} the high-dimensional feature space ($F$-d) into a much lower-dimensional one ($K$-d) within which the Euclidean distance is measured. Note that the low rank property reduces the number of model parameters and increases the training efficiency significantly.

\subsubsection{Multiple, Non-Metric Embeddings}
There are two key limitations from using a low-rank Mahalanobis embedding approach like the one above. First, it can capture only a single set of dimensions (or the `statistically dominant reason') that determines whether two given items are related or not. However, there might be multiple reasons relevant to the link discrimination task in question. For example, a shirt and a pair of pants might go well together due to complementary colors, compatible textures, or simply some common characteristics they share (such as both having pockets/buttons, etc.). This drives us to use a group of embeddings, parameterized by $N$ matrices $\mathbf{E}_1, \ldots, \mathbf{E}_N$ each with dimensionality $F \times K$ for the prediction task, with each capturing a different set of factors or `reasons' that items may be related. 

Another limitation of the single Mahalanobis embedding method, or more generally any metric-based method, is that it assumes that the closest neighbor of a given item is always itself, which is inappropriate for our task of placing many different categories of items close to the target. To overcome this shortcoming, we propose to use an anchor embedding (denoted by $\mathbf{E}_0$, again, with dimensionality $F \times K$) to learn the feature mappings in a non-metric manner.

In our model, $\mathbf{E}_0$ projects item $x$ to a \emph{reference point} $\mathbf{E}_0^T f_x$ in the corresponding space, referred to as the anchor space as it will be used as the basis for further comparisons. Next, embeddings $\mathbf{E}_k$ (for $k=1,2,\ldots,N$) map the potential match $y$ and correspond to a particular notion of relatedness, such that $\mathbf{E}_0^T f_x$ will be close to $\mathbf{E}_k^T f_y$ (for some $k$) if $x$ and $y$ are related.

That is, the predicted distance $d_k(x, y)$ by the $k$-th learner is
\begin{equation} \label{eq:dkxy}
d_k(x, y) = ||\overbrace{\mathbf{E}_0^T f_x}^{\mathclap{\text{$x$'s position in the anchor space}}} - \underbrace{\mathbf{E}_k^T f_y}_{\mathclap{\mathclap{\text{$y$'s position in the k-th `pseudo' space}}}}||_2^2.
\end{equation}
For clarity, we call the $N$ spaces defined by $\mathbf{E}_k$ ($k > 0$) `pseudo' spaces as all distance calculations are still performed within one actual space, i.e., the anchor space.

The above definition supports learning directed relationships as the model is not required to be symmetric; but, it is flexible enough to learn symmetric (or even metric) embeddings if such structures are exhibited by the data.

\subsubsection{Probabilistic Mixtures of Embeddings}

Now we introduce how we aggregate the predictions from different embeddings. Given an item pair $(x,y)$, we build our model upon the MoE framework to learn a probabilistic gating function to `switch' among different embeddings. Considering our asymmetric setting where the query item $x$ in the pair is used as the reference point, we model the probability that the $k$-th embedding is used for the given pair $(x, y)$ with a softmax formulation:
\begin{equation} \label{eq:gating}
P(k|\underbrace{(x, y)}_{\mathclap{\text{the given item pair}}}) = \overbrace{P(k|x)}^{\mathclap{\text{only depends on $x$}}} = \frac{\text{exp}(\mathbf{U}_{:,k}^T f_x)}{\sum_i \text{exp}(\mathbf{U}_{:,i}^T f_x)},
\end{equation}
where $\mathbf{U}$ is a newly-introduced $F \times N$ parameter matrix with $\mathbf{U}_{:,k}$ being its $k$-th column. Briefly, the idea is to compute the probability distribution over the $N$ learners given the characteristics of the `pivot' item $x$. Note that our formulation is efficient as it only introduces a small number of parameters given that $N$ is usually a small number (e.g.~on the order of 4 or 5 in our experiments).

Finally, our model calculates the `distance' of an item pair $(x,y)$ by the probabilistic expectation:
\begin{equation} \label{eq:dxy}
d(x, y) = \sum_k^N P(k|(x, y)) \cdot d_k(x, y).
\end{equation}

Note that our `distance' definition is a \emph{non-metric} method as it only preserves the non-negativity and is relaxing the symmetry, identity, and triangle inequality properties. 

\subsection{Learning the Model} \label{sec:learn}
With the `distance' function defined above, we model the probability that a pair is related by a shifted sigmoid function (in a way similar to \cite{VisualSIGIR}):
\begin{equation}
P((x,y)\in\mathcal{R}) = \sigma_c(-d(x,y)) = \frac{1}{1 + \exp(d(x,y) - c)}.
\end{equation}
Next, we need to randomly select a negative set of relationships $\mathcal{\bar{R}}$. To this end, we use a procedure from \cite{milo2003ugr} which randomly rewires the positive set in such a way that (1) the degree sequence of items is preserved and (2) each negative pair consists of items from two categories.

Then we proceed by fitting the parameters by maximizing the log-likelihood of the training corpus:
\begin{equation}
\begin{split}
\widehat{\Theta} = \arg\max_{\Theta} \mathcal{L}(\mathcal{R},\mathcal{\bar{R}} | \Theta) =\sum_{(x,y)\in\mathcal{R}} \log(P((x,y) \in \mathcal{R}))  \\
+\sum_{(x,y) \in \mathcal{\bar{R}}} \log(1 - P((x,y) \in \mathcal{R})) + \Omega(\Theta), 
\end{split}
\end{equation}
where $\Theta$ is the full parameter set $\{\mathbf{E}_0, \mathbf{E}_1, \ldots, \mathbf{E}_N, \mathbf{U}, c\}$, and $\Omega(\Theta)$ is an $\mathcal{L}_2$-regularizer to avoid overfitting. The total number of parameters is $F \times (N \times K + K + N) + 1$. Since $N$ and $K$ are {small} numbers (see Section \ref{sec:exp}), the log-likelihood as well as the derivatives can be computed efficiently.

\emph{\md} is learned with L-BFGS \cite{liu2009centroidal}, a quasi-Newton method for non-linear optimization of problems with a large number of variables. Our log-likelihood and the full derivative computations can be na{\"i}vely parallelized over all training pairs $(x,y) \in \mathcal{R} \cup \mathcal{\bar{R}}$. 
This means the optimization can easily benefit from multi-threading and even parallelization across multiple machines (e.g.~\cite{chen2014large}).
Note that \emph{\md} and the single-embedding method share the same time complexity when using the same amount of embedding parameters (see Appendix for a detailed analysis).

\section{Experiments} \label{sec:exp} 
\subsection{Dataset}
To fully evaluate the ability of \emph{\md{}} to handle real-world tasks, we want to experiment on the largest dataset available. To this end, we adopt the dataset from \emph{Amazon} recently introduced by \cite{VisualSIGIR}. We focus on five large top-level categories under the category tree rooted with `Clothing Shoes \& Jewelry', i.e., Men's, Women's, Boys', Girls', and Baby's Clothing \& Accessories. Statistics are shown in Table \ref{tb:clothingStat}.

For each of the above categories, we experiment with two important types of relationships: `users who bought \emph{x} also bought \emph{y},' and `users who viewed \emph{x} also viewed \emph{y},' denoted by `also\_bought' and `also\_viewed' respectively for brevity. Such relationships are a key source of data to learn from in order to recommend items of potential interests to customers. Ground-truth for these relationships is also introduced in \cite{VisualSIGIR}, and are originally derived from co-purchase and co-browsing data from \emph{Amazon}.

\begin{table}
\centering
\renewcommand{\tabcolsep}{4pt}
\caption{Statistics of a few representative categories from the \emph{Amazon} `Clothing Shoes \& Jewelry' dataset (using visual features).} 
\begin{tabular}{lrrrrrr} \toprule
\multirow{2}{*}{Dataset} &\multirow{2}{*}{\#Subcategories} &\multirow{2}{*}{\#Items} &\multicolumn{2}{c}{Relationship (\#Edges)} \\ \cline{4-5}
                         &             &          &\emph{also\_bought}  &\emph{also\_viewed} \\ \midrule
Men                      &56      	   &306,215   &1,075,547    &635,610     \\ 
Women                    &116      	   &659,566   &1,923,952    &1,691,121   \\
Boys                     &41	       &42,156    &169,503      &75,689      \\
Girls                    &44     	   &56,593    &191,964      &97,881      \\
Baby                     &6     	   &36,588    &96,253       &95,784      \\\midrule
Total                    &263          &1,101,118 &3,457,219    &2,596,085   \\ \bottomrule
\end{tabular}
\label{tb:clothingStat}
\end{table}

Recall that our objective is to learn heterogeneous relationships so as to support cross-category recommendation.
Across the entire dataset, such relationships are noisy, sparse, and not always meaningful. To address issues of noise and sparsity to some extent, it's sensible to focus on the relationships within the scope of a particular top-level category, e.g.~Women's Clothing, Men's Clothing etc.
We then consider relationships between `$2^{\text{nd}}$-level' categories, e.g.~women's shirts, women's shoes, etc.

In summary, our evaluation protocol is as follows:
\begin{enumerate}
\item A single experiment consists of a specific category (e.g. Men's Clothing) and a graph type (e.g. `also\_bought'). 
\item For each experiment, the relationships ($\mathcal{R}$) and a random sample of non-relationships 
($\bar{\mathcal{R}}$, see Section \ref{sec:learn})
are pairs of items connecting different subcategories of the category we are experimenting on. Note that $|\mathcal{R}| = |\bar{\mathcal{R}}|$ and they share the same distribution over the items.
\item For each experiment, we use an 80/10/10 random split of the dataset ($\mathcal{R} \cup \bar{\mathcal{R}}$) with the training set being at most two million pairs. Our goal is then to predict the relationships and non-relationships correctly, i.e., link prediction. 
\item For all methods, the \emph{validation} set is used for tuning the regularization hyperparameters, and finally the learned models are evaluated on the \emph{test} set in terms of error/misclassification rate.
\end{enumerate}

For example, one experiment is to predict `also\_bought' relationships for Men's Clothing. There are 56 subcategories under Men's Clothing (see Table \ref{tb:clothingStat}), so our goal is to distinguish edges from non-edges connecting items from among these subcategories.

All experiments were performed on a single machine with 64GB memory and 8 cores.
Our largest experiment required around 40 hours to train, though most were completed in a few hours.

\subsection{Comparison Methods} \label{sec:methods}

\xhdr{Weighted Nearest Neighbor (WNN):} This method uses a
weight\-ed Euclidean distance in the raw feature space to measure similarity between items: $d_w(x, y) = \| w \circ (f_x - f_y) \|^2_2$. Here $\circ$ is the Hadamard product and $w$ is a weighting vector that is learned from the data. 

\xhdr{Category Tree (CT):} This method computes a matrix of co-occur\-rences between subcategories from the training data. Then a pair $(x,y)$ is predicted to be positive if the subcategory of $y$ is one of the top 50\% most commonly connected subcategories to the subcategory of $x$. 

\xhdr{Low-rank Mahalanobis Transform (LMT):} LMT \cite{VisualSIGIR} is a state-of-the-art method for learning visual similarities among different items (possibly between categories) on large-scale datasets. LMT learns a \emph{single} low-rank Mahalanobis embedding matrix to embed all items into a low-dimensional space. Then it predicts the links between a given pair based on the Euclidean distance within the embedded space (i.e., \eq{eq:sigir}).

\xhdr{Mixtures of Non-metric Embeddings (\emph{\md{}}):} Our method. This method learns a mixture of low-rank transforms/embeddings to uncover groups of underlying reasons that explain the relationships between items. It measures the `distance' (or dissimilarity) between items in a non-metric manner (i.e., \eq{eq:dxy}).

Ultimately, our baselines are designed to demonstrate that (a) the raw feature space is not directly suitable for learning the notions of relationships (WNN); (b) using category metadata directly and not using other features (CT) results in relatively poor performance; 
and that (c) our proposed model is an improvement over the state-of-the-art method on our task (LMT).

\subsection{Performance \& Quantitative Analysis}
Error rates on the test set for all experiments are reported in Table \ref{tb:imgerr}. To perform a fair comparison between LMT and \emph{\md{}}, the following setting is used for all experiments in this paper:
\begin{enumerate}
\item It has been shown by \cite{VisualSIGIR} that LMT can achieve better accuracy when using a reasonably large number of embedding dimensions ($K$). 
Therefore in all cases we choose $K$ large enough such that LMT obtains the best possible (validation) performance.
\item In all cases we try to compare LMT and \emph{\md{}} under the same total number of model parameters. For example, if we set the number of dimensions $K$ to 100 for LMT, then a fair setting for \emph{\md{}} would be $K = 20$ and $N = 4$. This way both of them are using $100F$  embedding parameters. Recall that $N$ is the number of embeddings (excluding the anchor embedding).
\end{enumerate}

\begin{table}[t]
\centering
\renewcommand{\tabcolsep}{3pt}
\caption{Test errors of the link prediction task 
(i.e., predicting `also\_bought' and `also\_viewed' relationships between items) 
 using visual features (4096-d) for each edge type on clothing categories of the \emph{Amazon} dataset. The best performing method in each case is boldfaced. Lower is better.} 
\ \hspace{-20mm}
\begin{tabular}{lccccccc} \toprule
\multirow{2}{*}{Dataset} &\multirow{2}{*}{Graph}   &(a)     &(b)   &(c)  &(d)   &\% impr.   \\
                         &                    &WNN      &CT   &LMT     &\emph{\md{}}     & d vs.~c  \\ \midrule
\multirow{2}{*}{Men}     &\emph{also\_bought}  &34.95\%  &47.71\% &9.20\%  &\textbf{6.48\%}  &30\%  \\ 
						 &\emph{also\_viewed}  &18.98\%  &47.40\% &6.78\%  &\textbf{6.58\%}  &3\%   \\  [4pt]

\multirow{2}{*}{Women}   &\emph{also\_bought}  &30.50\%  &49.73\% &11.52\% &\textbf{7.87\%} &32\%  \\ 
						 &\emph{also\_viewed}  &20.50\%  &49.48\% &7.90\%  &\textbf{7.34\%} &7\%   \\  [4pt]
                         
\multirow{2}{*}{Boys}    &\emph{also\_bought}  &31.16\%  &46.02\% &8.80\%  &\textbf{5.71\%} &35\%  \\ 
						 &\emph{also\_viewed}  &21.52\%  &46.22\% &6.72\%  &\textbf{5.35\%} &20\%  \\  [4pt]

\multirow{2}{*}{Girls}   &\emph{also\_bought}  &31.10\%  &47.63\% &8.33\%  &\textbf{5.78\%} &31\%  \\ 
						 &\emph{also\_viewed}  &22.36\%  &46.43\% &6.46\%  &\textbf{5.62\%} &13\%  \\  [4pt]
                         
\multirow{2}{*}{Baby}    &\emph{also\_bought}  &37.26\%  &48.01\% &12.48\% &\textbf{7.94\%} &36\%  \\ 
						 &\emph{also\_viewed}  &30.89\%  &47.72\% &11.88\% &\textbf{9.25\%} &22\%  \\ \midrule
Avg.                     &                    &27.92\%  &47.64\% &9.00\%  &\textbf{6.79\%} &22.9\%  \\ \bottomrule
\end{tabular}
\hspace{-20mm}\
\label{tb:imgerr}
\end{table}

For experiments on `also\_bought' relationships, LMT uses $K = 100$ dimensions and \emph{\md{}} uses $K=20$ and $N=4$. While for experiments on `also\_viewed' relationships, $K$ is set to $50$ for LMT and $K=10$ and $N=4$ for \emph{\md{}}. Note that `also\_viewed' relationships are almost twice as sparse as `also\_bought' relationships (and thus a model with fewer parameters performed better at validation time), as shown in Table \ref{tb:clothingStat}. We make a few observations to explain and understand our findings as follows:

\begin{figure*}
\centering
\includegraphics[width=\linewidth]{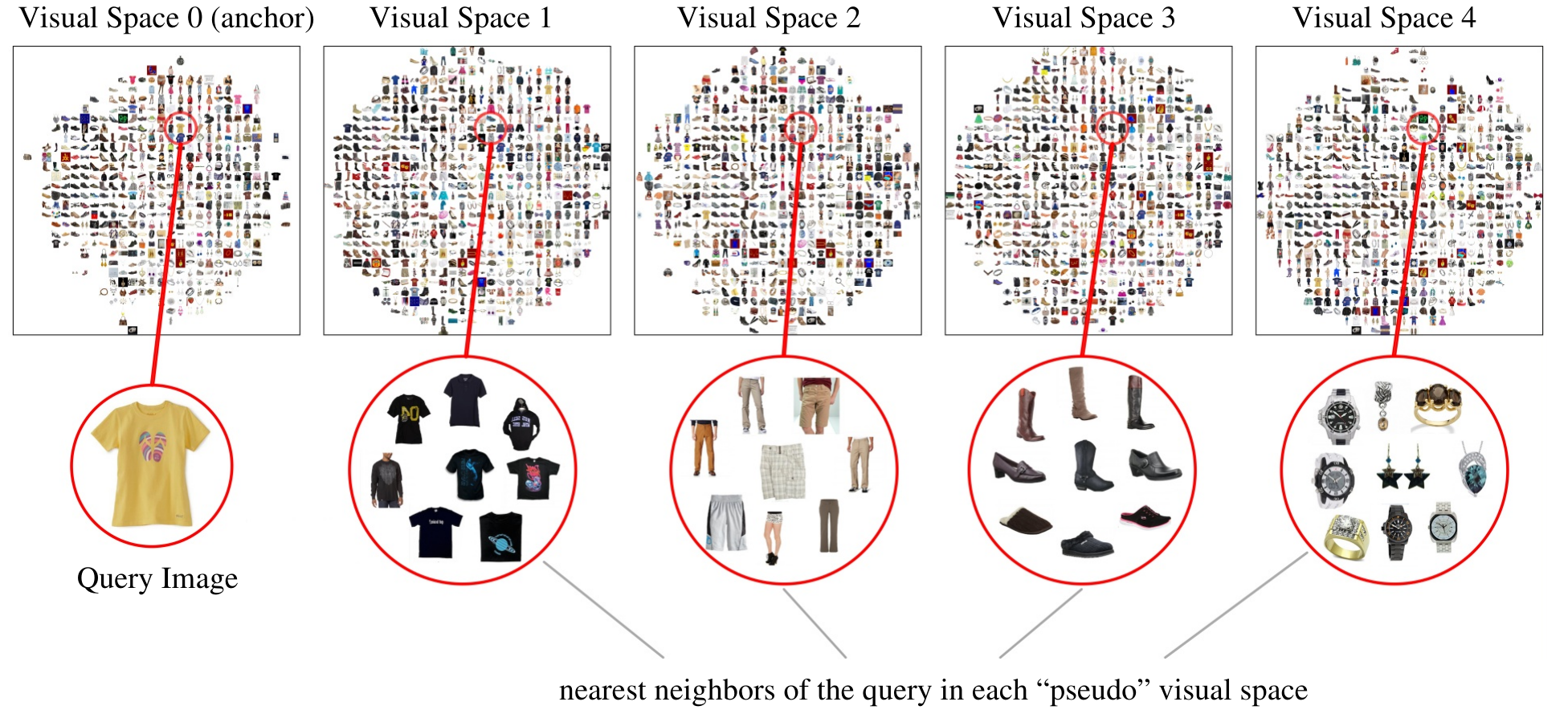}
\caption[Caption without FN]{Visualization of \emph{\md{}} trained on Women's Clothing for `also\_bought' prediction. Each visual space is demonstrated by a 2-d t-SNE grid view \cite{tsne} (each cell randomly selects one image in overlapping cases). According to our `distance' function (i.e., \eq{eq:dxy}), \emph{\md{}} recommends the nearest neighbors of the query within each visual space, based on the associated `reasons' learned from data.
Note that each visual space exhibits different category `clusters' at the query image's location, allowing us to recommend {diverse} sets of items from the most closely-related categories.}
\label{fig:todo}
\end{figure*}

\begin{enumerate}
\item WNN is particularly inaccurate for our task. We also observed relatively high training errors with this method for most experiments. This confirms our conjecture that raw similarity is inappropriate for our task, and that in order to learn the relationships across (sub)categories, some sort of expressive transforms are needed for manipulating the raw features.

\item The counting method (CT) performs considerably worse than other methods. This reveals that the predictive information used by the other models goes beyond the categories of the products, i.e., that the image-based models are learning relationships between finer-grained attributes.

\item Note that all models perform better at predicting `also\_viewed' than `also\_bought' relationships. This is reasonable since intuitively items that are ``also viewed'' indeed tend to share more common characteristics compared to the ``also bought'' scenario. The greater heterogeneity between training pairs in the latter task makes it comparatively harder to address.

\item \emph{\md{}} outperforms LMT significantly for all experiments, especially for the harder task of predicting co-purchase dyads. 
\end{enumerate}

\subsection{Visualization of the Embeddings}
Next, we proceed by demonstrating the embeddings learned from our largest dataset, Women's Clothing, by \emph{\md{}}. We take the same model trained on co-purchase relationships from the previous subsection and visualize it in Figure~\ref{fig:todo}. In this figure, we show each of the 5 visual spaces by a 2-d visualization with t-SNE \cite{tsne}. Images are a random sample of size 50,000 from the Women's Clothing dataset and projected (using the learned embedding matrices) to each visual space to demonstrate the underlying structure.

As analyzed in Section \ref{sec:model}, each embedding (i.e., learner) is capturing a specific notion of relatedness that explains the relationships of pairs of items in the corpus. In other words, it means that the nearest neighbors in each of the $N$ `pseudo' spaces should be related to the query according to the specific notion captured. Therefore those neighbors should be recommended as potential matches to the query item, as shown by the example in Figure~\ref{fig:todo}. For the query image (a t-shirt) in this example, \emph{\md{}} recommends bundles of similar t-shirts, pants, shoes, and accessories (watches etc.) that resemble the query in terms of patterns (e.g.~space 1), colors (e.g.~space 2), and more generally `styles' (e.g.~space 3 and 4).\footnote{The second patch actually contains a few men's clothing items due to data deficiency---an intrinsic problem suffered by \emph{Amazon}.}
Such matching between a query image and nearby items in alternate spaces directly facilitates the task of recommending visually consistent outfits, where modeling and understanding the visual compatibility across categories is essential.

\begin{figure} 
\begin{center}
\includegraphics[width=\linewidth]{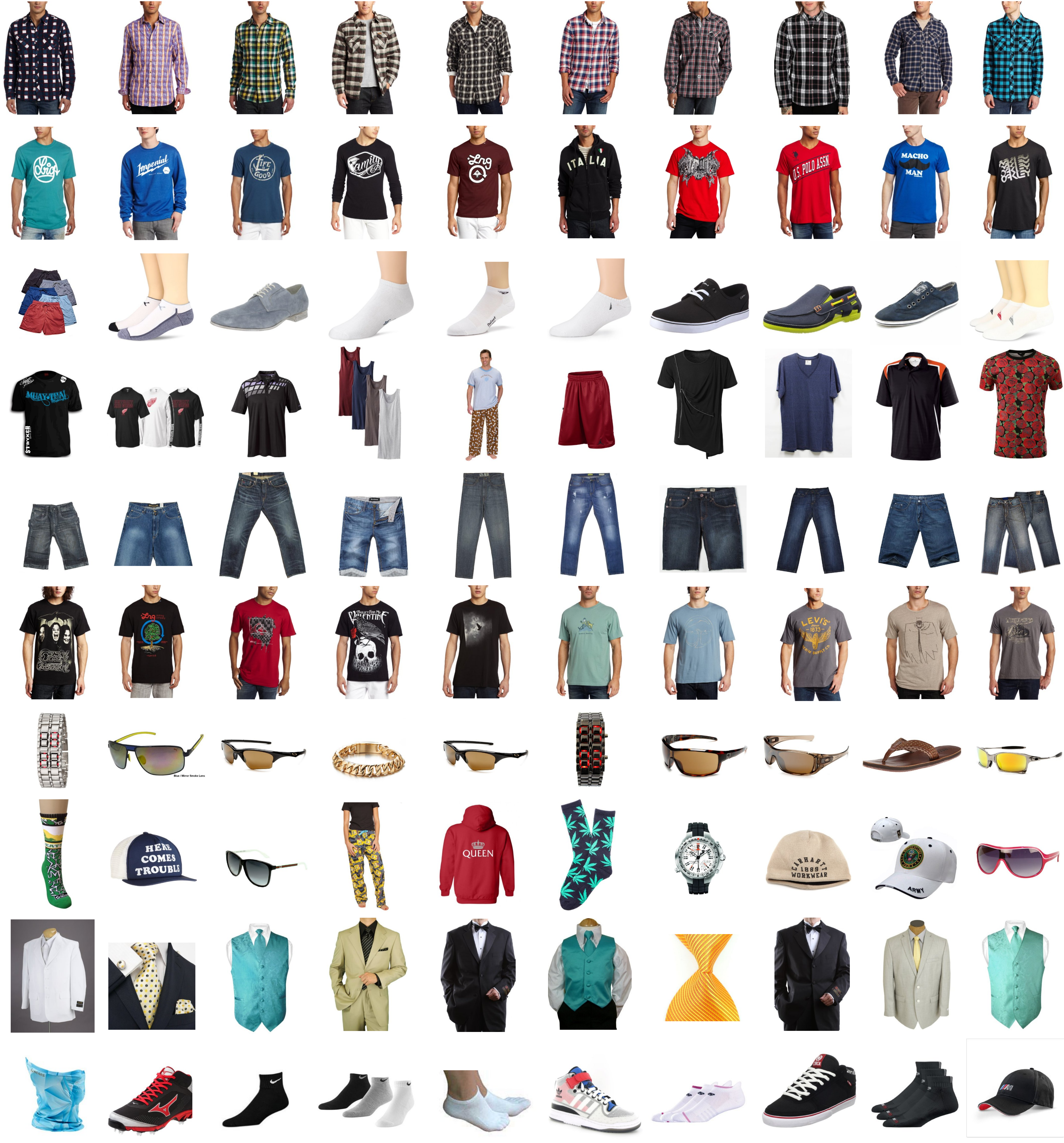}
\end{center}
\caption{Demonstration of the 10 visual dimensions of one (randomly selected) visual space learned by \emph{\md{}} on Men's Clothing for co-purchase prediction ($K=10, N=4$). Each row shows the top ranked items for a particular dimension $i$, i.e., $\arg\max_x \mathbf{E}_{:,i}^T f_x$ where $\mathbf{E}_{:,i}$ is the $i$-th column of the corresponding embedding matrix $\mathbf{E}$.}
\label{fig:dims}
\end{figure}

\subsection{Visualization of the Visual Dimensions}
Next we demonstrate the visual dimensions learned by \emph{\md{}}, i.e., what kind of characteristics the model is capturing to explain the relationships among items. A simple way to visualize these dimensions is to show items that exhibit maximal values for each dimension. In other words, we select items according to 
$$\arg\max_x \mathbf{E}_{:,i}^T f_x,$$
where $\mathbf{E}_{:,i}$ is the $i$-th column of the embedding matrix $\mathbf{E}$, corresponding to a visual dimension $i$.
Intuitively, this informs us of items that are most representative of a particular visual aspect discovered by the model.

We trained \emph{\md{}} on Men's Clothing ($K=10, N = 4$), predicting co-purchase relationships. Due to limited space, we randomly select one embedding and demonstrate its 10 visual dimensions in Figure~\ref{fig:dims}. 
From the figure we can see that (1) \emph{\md{}} seems to uncover meaningful visual dimensions, each of which highlights certain \emph{fine-grained} item types (e.g.~plaid tees and jeans in row 1 and 5); (2) human notions seem to have been captured, e.g.~casual versus formal in rows 2 and 9; and (3) subtle differences between different characteristics can be distinguished (e.g.~tees in rows 2 and 6). \emph{\md{}}'s ability to discover and model the correlations among visual characteristics explains its success.

\subsection{Visual Recommendation \& Analysis}
Beyond achieving high prediction accuracy, we want to test the ability of \emph{\md{}} to generate useful recommendations. Again, we mainly compare to the state-of-the-art metric-based method, LMT. Both methods are able to learn relationships from the data, so one common setting is to retrieve `similar' items (i.e., maximum probability of being related) to a given query.

\begin{figure}[t] 
\centering
\includegraphics[width=\columnwidth]{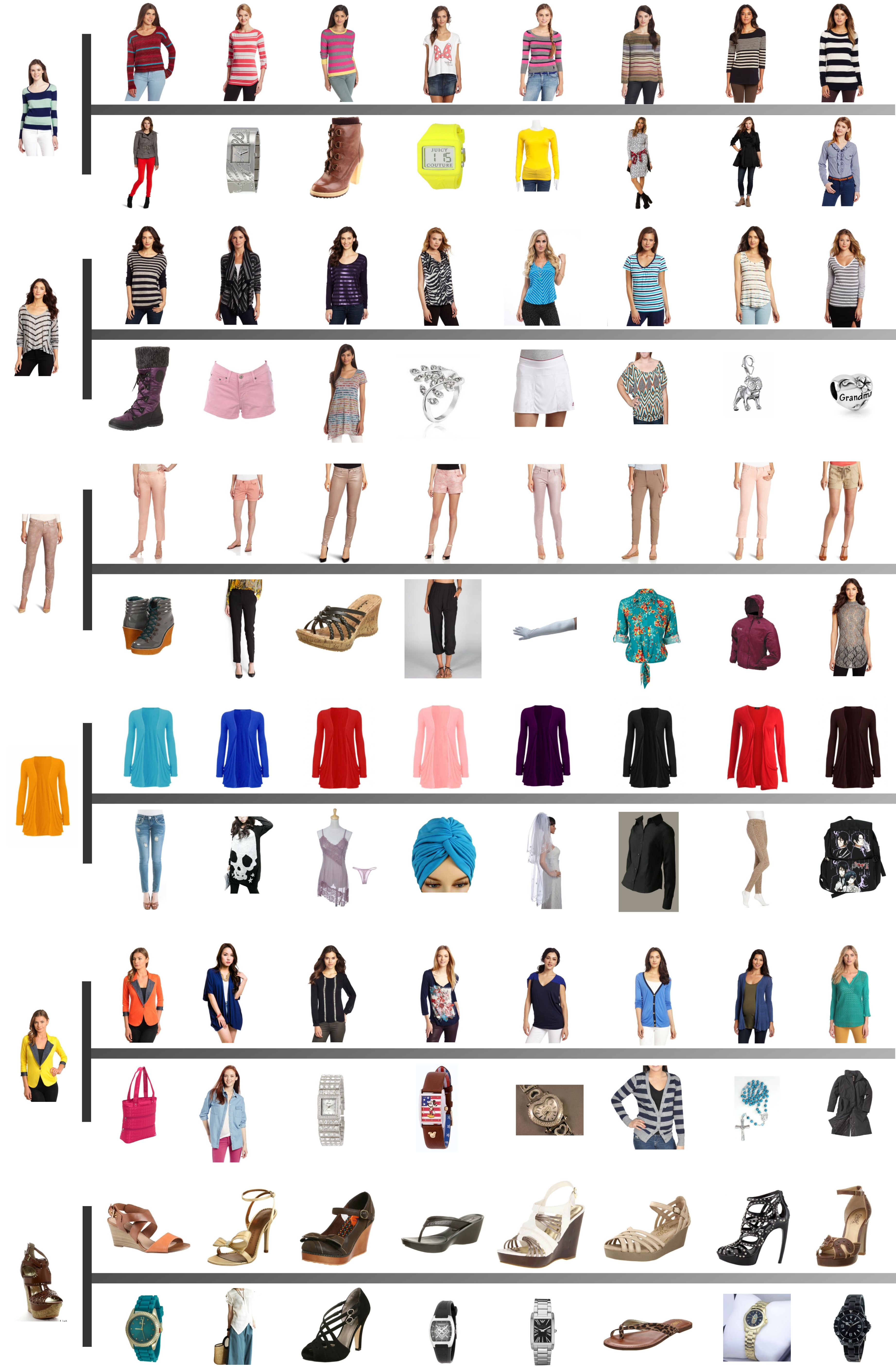}
\caption{Comparison between the state-of-the-art metric-based method, LMT, and our non-metric method \emph{\md{}}. On the left are a few query images, for each of which we show its nearest neighbors retrieved by LMT (above the horizontal line) and \emph{\md{}} (below the horizontal line) respectively. Both models were trained on Women's Clothing for `also\_bought' prediction (using the same setting as in Table \ref{tb:imgerr}). All query images and all neighbors are from Women's Clothing.}
\label{fig:nn}
\end{figure}
\setlength{\textfloatsep}{13pt plus 2pt minus 4pt}

First we train LMT and \emph{\md{}} on Women's Clothing to predict `also\_bought' relationships, under the same setting as in Table \ref{tb:imgerr}. This way the two models will learn their own similarity (or distance) functions from the data. Next, from Women's Clothing we randomly select a few query items, for each of which LMT and \emph{\md{}} will retrieve its highest-probability links according to their own similarity functions. Figure~\ref{fig:nn} demonstrates such queries and the retrieved connections (in all cases ranked in decreasing order in terms of the probability of the link) by the two models.

As shown in Figure~\ref{fig:nn}, the metric-based method (LMT) tends to recommend items that are very similar to the query, even though for this task it is trained to predict complementary relationships (i.e., `also\_bought'). Indeed it is very difficult for a metric-based method to project items from different subcategories to be nearer than items from the same category; presumably such methods are limited by their underlying assumption that the most similar item to a given query is always itself. In \cite{VisualSIGIR} this was addressed to some extent by making explicit use of the category information at test time (e.g.~`find the shirt closest to this pair of shoes'), though our model is able to make diverse sets of recommendations without such a dependence on explicit category information.

\begin{figure}
\centering
\begin{subfigure}{.24\textwidth}
  \centering
  \includegraphics[width=.96\linewidth]{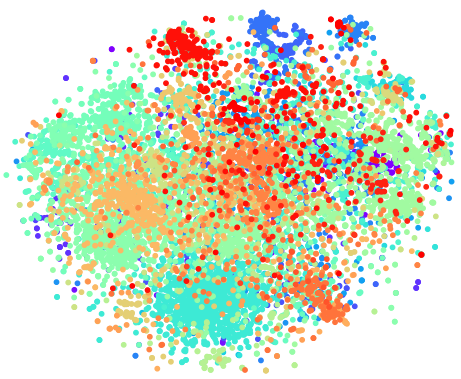}
  \caption{Men's Clothing}
  \label{fig:submen}
\end{subfigure}%
\begin{subfigure}{.24\textwidth}
  \centering
  \includegraphics[width=.96\linewidth]{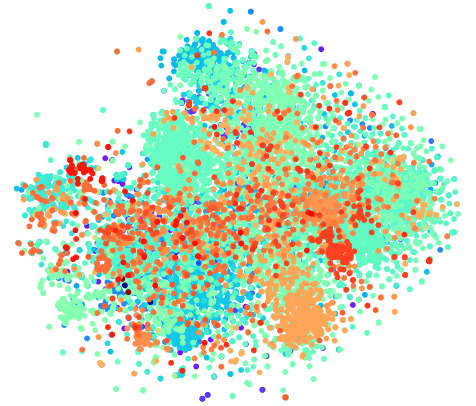}
  \caption{Women's Clothing}
  \label{fig:subwomen}
\end{subfigure}
\caption{Demonstration of the distribution of different subcategories in the 100-d space learned by LMT. For visualization, we use t-SNE~\cite{tsne} to further embed this space into 2-d. In each subgraph, a color represents a specific subcategory within the corresponding dataset. The main finding is that LMT tends to project subcategories to be `clusters' in the embedded space, which may cause a `limited coverage' problem for the recommendation task.}
\label{fig:lmtembed}
\end{figure}

Recall that LMT learns an embedding within which the Euclidean distance is used to distinguish relationships from non-relationships. Visualizing such spaces can help understand the behavior of LMT. Again we uniformly sample 10,000 items from the dataset and use t-SNE \cite{tsne} to visualize their positions in the embedded space. We are particularly interested in the distribution of different subcategories of items over the space. Therefore we assign a unique color to each subcategory in the dataset. Figure~\ref{fig:lmtembed} shows results on two representative datasets, Men's and Women's Clothing.

From Figure~\ref{fig:lmtembed} we find that subcategories of items tend to become `clusters' in the embedded space. This can be problematic especially for recommending related items across subcategories:
\begin{enumerate}
\item From a recommendation perspective, there will be a `limited coverage' issue because given a query LMT tends to recommend \emph{only} those items on the \emph{boundaries} of the clusters. 
There is no way that items located near the center of a cluster will ever be recommended, since the closest item outside the query's own category cluster will be on the fringe of a different category cluster.
\item Recommendations will suffer from `mislabeling' issues. Note that LMT relies heavily on the taxonomy metadata at test time to filter out items from the same subcategory as the query. However, even a tiny number of mislabeled items in the dataset can poison recommendations, and certainly some such examples exist in the \emph{Amazon} dataset \cite{SiameseICCV}. Unexpected items may appear on the recommendation list when there are mislabeled items that actually come from the same subcategory as the query.
\end{enumerate}

\begin{table}[t]
\centering
\renewcommand{\tabcolsep}{6pt}
\caption{Statistics of a variety of top-level categories from \emph{Amazon}.} 
\begin{tabular}{lrrrrrrr} \toprule
\multirow{2}{*}{Dataset} &\multirow{2}{*}{\#Subcategories} &\multirow{2}{*}{\#Items} &\multicolumn{2}{c}{Relationship (\#Edges)} \\ \cline{4-5}
        &           &          &\emph{also\_bought}  &\emph{also\_viewed} \\ \midrule
Elect   & 306       &412,082   &1,654,552 &718,361     \\
Auto    & 178       &312,642   &959,353   &1,298,774   \\ 
Games   & 16        &49,801    &314,124   &54,559      \\ 
Movies  &  2        &199,737   &648,256   &49,924      \\
Office  &245        &127,054   &448,720   &370,630     \\ 
Home    & 81        &393,781   &560,574   &960,925     \\ 
Phones  & 28        &317,965   &867,418   &225,785     \\ \midrule
Total   & 856       &1,813,062 &5,452,997 &3,678,958   \\ \bottomrule
\end{tabular}
\label{tb:more}
\end{table}

Note that \emph{\md{}} doesn't suffer from either of above issues since it has already successfully `blended' different subcategories, as shown by the nearest neighbors in Figure~\ref{fig:nn}.

\subsection{Learning Compatibility from Textual Features} \label{sec:generalize}
In previous sections, we have shown that \emph{\md{}} not only performs very well on link prediction tasks but also that it recommends highly diverse sets of items. However, above we only considered scenarios in which relationships like co-purchasing can be predicted from visual features. Following this, a natural question would be ``Is \emph{\md{}} able to learn relationships from \emph{non}-visual features and achieve similarly competitive performance?'' 

To answer the above question, we perform further experiments on Bag-of-Words (BoW) features extracted from the text of product reviews, which are also available in the \emph{Amazon} dataset. In particular, we experiment with a variety of top-level \emph{Amazon} categories, i.e, Electronics, Automotive, Video Games, Movies \& TV, Office Products, Home \& Kitchen, and Cell Phones \& Accessories (denoted by `Elect', `Auto', `Games', `Movies', `Office', `Home', and `Phones' resp.). Statistics of these datasets are shown in Table \ref{tb:more}.

For each category (e.g.~Electronics) we use the following procedure to generate BoW features for all its items: (1) Remove stop-words and construct a dictionary. Our dictionary consists of 5000 nouns or adjectives or adjective-noun bigrams that appear most frequently in the review corpus being considered. (2) For each item $i$, a document $d_i$ is generated by bagging all the reviews it has received. (3) The 5000-d BoW feature vector $f_i$ of each item $i$ is computed by normalizing the raw word counts of document $d_i$ to sum to 1. (4) Items without any reviews attached are seen as invalid items and are dropped from the dataset.
In the following experiments, we use the same evaluation protocol as in the previous visual feature experiments.

\xhdr{Latent Dirichlet Allocation + WNN (LDA):} Here we add another baseline for further comparison. This method first obtains 100 topics with LDA with a vocabulary of size 5000,\footnote{We adopted the implementation in Gensim (default parameters kept): https://radimrehurek.com/gensim/} and then uses WNN to distinguish relationships from non-relationships within the 100-d topic space.

\xhdr{Results and Analysis.}
Table \ref{tb:bowerr2} summarizes the error rates on the test sets for all experiments. We observe that (1) basic methods like WNN and LDA are not particularly accurate for the task; (2) \emph{\md{}} outperforms LMT considerably especially on the harder tasks, which demonstrates its ability to handle textual features; and (3) the comparative hardness of `also\_bought' over `also\_viewed' prediction now seems to be dependent on the dataset in question, presumably due to different semantics of the two link types, or different patterns of customer behavior, among different categories.

\begin{table} [!t]
\centering
\renewcommand{\tabcolsep}{4pt}
\caption{Test errors of the link prediction task using BoW features (5000-d) on a variety of top-level categories of the \emph{Amazon} dataset. For LMT, $K = 100$, while for \emph{\md{}}, $K = 20$ and $N = 4$. Lower is better.} 
\ \hspace{-20mm}
\begin{tabular}{lccccccc} \toprule
\multirow{2}{*}{Dataset} &\multirow{2}{*}{Graph}   &(a)   &(b)   &(c)   &(d)   &\% impr.  \\
                         &                    &WNN   &LDA   &LMT   &\emph{\md{}}   & d vs.~c \\ \midrule
\multirow{2}{*}{Elect}   &\emph{also\_bought}  &37.58\% &36.67\%  &13.73\%  &\textbf{10.09\%} &26\%  \\ 
						 &\emph{also\_viewed}  &39.37\% &26.60\%  &16.41\%  &\textbf{9.44\%}  &42\%  \\  [4pt]

\multirow{2}{*}{Auto}    &\emph{also\_bought}  &42.63\% &38.57\%  &17.94\%  &\textbf{14.09\%} &21\%  \\ 
						 &\emph{also\_viewed}  &42.44\% &34.37\%  &21.15\%  &\textbf{14.74\%} &30\%  \\  [4pt]
                         
\multirow{2}{*}{Games}   &\emph{also\_bought}  &44.31\% &42.55\% &14.43\%  &\textbf{12.03\%} &17\%  \\ 
						 &\emph{also\_viewed}  &40.08\% &33.22\% &16.18\%  &\textbf{11.29\%} &30\%  \\  [4pt]

\multirow{2}{*}{Movies}  &\emph{also\_bought}  &40.00\% &23.51\% &11.36\%  &\textbf{9.47\%}  &17\% \\ 
						 &\emph{also\_viewed}  &42.01\% &26.63\% &15.12\%  &\textbf{14.98\%} &1\%  \\  [4pt]

\multirow{2}{*}{Office}  &\emph{also\_bought}  &41.35\% &39.05\% &18.53\%  &\textbf{14.30\%} &23\% \\ 
						 &\emph{also\_viewed}  &37.33\% &28.13\% &13.52\%  &\textbf{9.72\%} &28\%  \\  [4pt]

\multirow{2}{*}{Home}    &\emph{also\_bought}  &39.74\% &31.91\%  &13.96\%  &\textbf{12.40\%} &11\%  \\ 
						 &\emph{also\_viewed}  &36.49\% &23.09\%  &13.97\%  &\textbf{9.95\%}  &29\%  \\  [4pt]
                         
\multirow{2}{*}{Phones}  &\emph{also\_bought}  &43.35\% &42.73\%  &29.44\%  &\textbf{22.68\%} &23\%  \\ 
						 &\emph{also\_viewed}  &43.70\% &34.30\%  &24.65\%  &\textbf{16.04\%} &35\%  \\   \midrule
Avg.                     &                    &40.74\% &32.95\%  &17.19\%  &\textbf{12.94\%} &23.8\%  \\ \bottomrule
\end{tabular}
\hspace{-20mm}\ 
\label{tb:bowerr2}
\end{table}

\section{Conclusion}
In this paper, we presented \emph{\md{}}, a method to model heterogeneous relationships for item-to-item recommendation tasks. We noted that existing methods for item-to-item recommendation suffer from a few limitations when dealing with heterogeneous data, due mainly to their reliance on metricity or `nearest-neighbor' type assumptions. To overcome these limitations, our method made use of `mixtures' of non-metric embeddings, which allows us to relax the identity and symmetry assumptions of existing metric-based methods. The proposed 
scalable 
approach generates diverse and cross-category recommendations effectively that capture more complex relationships than mere visual similarity. We showed quantitatively that \emph{\md{}} is accurate at link prediction tasks using co-purchase and co-browsing dyads from \emph{Amazon}, and qualitatively that it is able to generate diverse recommendations that are consistent with a particular visual style.

\appendix
\xhdr{Complexity Analysis:} In \emph{\md{}}, each embedding matrix has $F \times K$ parameters, which means there are in total $F \times K \times (N + 1)$ embedding parameters. Since \emph{\md{}} doesn't need to use more embedding parameters to outperform LMT (see Section \ref{sec:exp}), we focus on comparing \emph{\md{}} and LMT under the same total number of embedding parameters, in terms of the amount of multiplications involved.

For complete clarity, we denote the embedding dimension of LMT and \emph{\md{}} by $K'$ and $K$ respectively ($F \times K' = F \times K \times (N + 1)$). For each training pair $(x,y)$, LMT takes $\mathcal{O}(F \times K')$ to compute the distance between them and the corresponding derivatives. While for \emph{\md{}}, it takes $\mathcal{O}(F \times K')$ to project $x$ and $y$ to the multiple spaces. Afterwards the `distance' will be calculated in $\mathcal{O}(N \times K) + \mathcal{O}(N \times F)$, where the former is for computing $N$ distance components and the latter is spent on the probabilistic weights. In total, it takes $\mathcal{O}(F \times K') + \mathcal{O}(N \times K) + \mathcal{O}(N \times F) = \mathcal{O}(F \times K')$ for \emph{\md{}} to finish `distance' computation. Likewise, it's easy to verify that the corresponding derivatives can also be computed in $\mathcal{O}(F \times K')$ time. 
To sum up, training \emph{\md{}} and LMT will have the same time complexity when using the same amount of embedding parameters.

\balance
\bibliographystyle{abbrv}
\bibliography{sigproc}

\end{document}